\documentclass[pre,aps,superscriptaddress,preprint]{revtex4-1}

\usepackage{amssymb,amsmath}
\usepackage{graphicx}
\usepackage{nicefrac}
\usepackage{color}
\usepackage{upgreek}

\usepackage{soul} 
\usepackage{dcolumn}
\usepackage{bm}
\newcommand\p[2]{\frac{\partial #1}{\partial #2}}
\newcommand\pp[3]{\frac{\partial ^{#3} #1}{\partial #2 ^ {#3}}}
\newcommand\ti[1]{\widetilde{#1}}

\begin{document}

\title
{Self-similar solution of the subsonic radiative heat equations using a binary equation of state}
\author
{Shay I. Heizler}
\email{highzlers@walla.co.il}
\affiliation{Department of Physics, Bar-Ilan University, Ramat-Gan, IL52900 ISRAEL}
\affiliation{Department of Physics, Nuclear Research Center-Negev, P.O. Box 9001, Beer Sheva 84190, ISRAEL}
\author
{Tomer Shussman}
\email{tomer.shussman@mail.tau.ac.il}
\affiliation{Raymond and Beverly Sackler School of Physics \& Astronomy, Tel Aviv University, Tel Aviv 69978, ISRAEL}
\affiliation{Department of Plasma Physics, Soreq Nuclear Research Center, Yavne 81800, ISRAEL}
\author
{Elad Malka}
\affiliation{Israel Atomic Energy Commission, P.O. Box 7061, Tel Aviv 61070, ISRAEL}

\begin{abstract}
Radiative subsonic heat waves, and their radiation driven shock waves, are important hydro-radiative phenomena.
The high pressure, causes hot matter in the rear part of the heat wave to ablate backwards. At the front of the heat wave,
this ablation pressure generates a shock wave which propagates ahead of the heat front. Although no self-similar solution of both the ablation and shock regions exists,
a solution for the full problem was found in a previous work. Here, we use this model in order to investigate
the effect of the equation of state (EOS) on the propagation of radiation driven shocks. We find that using a single ideal
gas EOS for both regions, as used in previous works, yields large errors in describing the shock wave.
We use the fact that the solution is composed of two different self-similar solutions,
one for the ablation region and one for the shock, and apply two ideal gas EOS (binary-EOS), one for each region, by fitting a detailed tabulated EOS to power laws at different regimes.
By comparing the semi-analytic solution with a numerical simulation using
a full EOS, we find that the semi-analytic solution describes both the heat and the shock regions well.

\end{abstract}

\maketitle
 
\section{Introduction}

Over the last few decades there is a growing interest in radiative heat-waves, specifically for High Energy Density Physics (HEDP) experiments~\cite{lindl,rosen}.
In these experiments, a source of X-ray heats the boundary of a target, and a radiative heat wave propagates through the target's material.
If the problem contains more than a few mean free paths, local thermodynamic equilibrium (LTE) and diffusion approximation can be assumed.
When the heat-wave propagation is faster than the speed of sound (supersonic case), it is described using only the radiative transfer equation~\cite{zeldovich,pombook,heizler}.

When the heat front propagates slower than the speed of sound (subsonic case), hydrodynamic motion of the matter becomes important.
The high pressure of the heated matter causes ablation behind the heat front, thus the density is low near the rear boundary,
and high near the ablation front. This ablation pressure generates a strong shock wave which propagates from the heat front forward.
We call the former region ``ablation region" and the latter ``shock region".
Subsonic heat waves are important in modeling inertial confinement fusion (ICF) indirect-drive experiments~\cite{lindl},
specifically in the interior of the high-Z optically thick hohlraum walls.

The heat wave problem was first stated by Marshak~\cite{marshak}, who proposed self-similar solutions for the supersonic case. Subsonic case solutions were investigated by many authors~\cite{ps,ps2,ger3,hr,garnier}. Most of the solutions focused on the ablation region alone, since the full problem (including the shock region) is not self-similar for the general case. Early simple approximations offered taking a constant
ablation pressure in the shock regime~\cite{tlusty}. However, these approximations may yield large errors, especially if the yielding ablation pressure is far from being constant, such as the case of constant temperature boundary condition.

Recently, Shussman and Heizler proposed a semi-analytic model to describe both ablation and shock regions~\cite{ts}.
The solution is composed of two different self-similar solutions, one for each region. The two solutions are patched together at the heat front. The model was compared to full numerical simulations (which used the same single ideal gas EOS) of subsonic radiative heat wave in Au, and found to fit the simulations to within $1\%$ in the heat region and $3\%$ in the shock region.

However, the semi-analytic model (and the simulations) used the same EOS for both the ablation and shock regions. Using a single ideal-gas for the EOS (which is fitted for the ablation regime as in~\cite{hr}) may cause large errors when describing the shock region, since the temperatures and densities vary widely between the ablation region and the shock region. This can be shown by comparing
the semi-analytic solution to a numerical simulation with an EOS tables.

In this work, we propose a correction for the semi-analytic model to well describe the shock region.
We use the fact that the solution is composed of two self-similar solutions, and therefore there is no
need to use only one EOS for both solutions. In order to conserve the self-similar nature of each solution,
we assume ideal gas EOS, but choose a different ideal gas parameter $\gamma\equiv r+1$ for each region. This method
is therefore equivalent to using a binary-EOS with piecewise temperature dependency. We examine this differentiation
caused by using the binary-EOS to an exact SESAME table~\cite{sesame}. 

This paper proceeds as follows: In Sec. \ref{model} we briefly present the model and the governing equations, which were introduced widely in~\cite{ts}, only now using a binary-EOS.
In Sec. \ref{motivation} we present the results of simulations using a single EOS, compared to using a SESAME table EOS, and show that large errors are obtained by using a single EOS for both regions. Thus, we stress the importance of using a different EOS for each region. In Sec. \ref{results} we investigate the binary-EOS model, and compare the results to numerical calculations using binary-value EOS or SESAME table EOS. A short discussion is presented in Sec. \ref{discussion}.

\section{Model and main methodology}
\label{model}

In the section we briefly repeat the methodology of obtaining and patching two self-similar solutions in order to describe the whole solution (for details see ~\cite{ts}). The ablation region is solved using a time-dependent temperature boundary condition, in a manner similar to ~\cite{ps}, and the shock region is solved using a time-dependent pressure boundary condition. The pressure at the ablation front (ablation pressure) serves as the boundary condition for the shock region.

\subsection{The ablation region}

For the subsonic case, the hydro-radiation set of equations under LTE and diffusion approximations are:  
\begin{subequations}
\label{sub_basic}
\begin{equation}
\label{mass}
\p{V}{t}-\p{u}{m}=0
\end{equation}
\begin{equation}
\label{momentum}
\p{u}{t}-\p{P}{m}=0
\end{equation}
\begin{equation}
\label{sub_energy}
\p{e}{t}+P\p{V}{t}=\p{}{m}\left(\frac{c}{3\kappa_R}\p{\left(aT^4\right)}{m}\right)
\end{equation}
\end{subequations}
Here, $P$ is the pressure, $V$ is the specific volume ($V=1/\rho$ where $\rho$ is the density of the material), $u$ is matter velocity and $m$, the Lagrangian
coordinate, is defined by $m(x,t)=\int_0^x \rho(x',t)dx'$ ($x$ denotes the Eulerian coordinate and $t$ is time). $a\equiv\nicefrac{4\sigma}{c}$ is the radiation
density constant ($\sigma$ is Stefan-Boltzmann constant) and $c$ is the speed of light.
In order to find a self-similar solution, we assume the internal energy of the matter $e$ and the opacity $\kappa_R$ can be expressed in the form of power
laws of the density and the temperature $T$ (we adopt the notation first presented by~\cite{hr}):
\begin{subequations} \label{pwrlaws}
\begin{equation} \label{ross}
\frac{1}{\kappa_R}=gT^\alpha\rho^{-\lambda}
\end{equation}
\begin{equation} \label{pwrlaw_energy}
e=fT^\beta\rho^{-\mu}
\end{equation}
\end{subequations}

In addition, we assume the EOS in the ablation region is well approximated by an ideal gas,
\begin{equation}
P(\rho,T)=r_1\rho e(\rho,T)\equiv(\gamma_1-1)\rho e(\rho,T)
\label{def_r}
\end{equation}
where $r_1\equiv(\gamma_1-1)$ is the ideal gas parameter in the ablation region.

Using Eqs. \ref{pwrlaw_energy} and \ref{def_r}, the temperature $T$ can be expressed as a function of $P$ and $V$:
\begin{equation}
T=\left(\frac{PV^{1-\mu}}{r_1f}\right)^{\frac{1}{\beta}}
\label{def_T}
\end{equation}

The numerical values for the internal energy and the opacity in the heat wave region, which are best fit for a temperature range of $100-300\mathrm{eV}$
and are taken from~\cite{hr}, are presented in Table \ref{table:pwr_law_opac_eos}.
\begin{table}[!htb]
  \centering
  \caption{\bf Power law fits for the opacity and EOS of Au in temperatures $1-3\mathrm{HeV}$~\cite{hr}}
  \label{table:pwr_law_opac_eos} 
  \begin{tabular}{|c|c|} \hline 
    \multicolumn{1}{|c|}{{\bf Physical Quantity}} &
    \multicolumn{1}{c|}{{\bf Numerical Value}} \\ \hline
    \ $f$ & $3.4$ [MJ/g]  \\ \hline
    \ $\beta$ & $1.6$  \\ \hline
    \ $\mu$ & $0.14$  \\ \hline
    \ $g$ & $1/7200$ $[\mathrm{g/cm^2}]$  \\ \hline
    \ $\alpha$ & $1.5$  \\ \hline
    \ $\lambda$ & $0.2$  \\ \hline
    \ $r_1\equiv(\gamma_1-1)$ & $0.25$  \\ \hline
  \end{tabular}
\end{table}

We also assume that the boundary condition has a power-law form:
\begin{equation} \label{temperature_pwrlaw}
T(t)=T_0t^{\uptau}
\end{equation}

After substituting Eqs. \ref{def_T}-\ref{temperature_pwrlaw} into Eq. \ref{sub_energy}, it can be seen that the three Eqs. \ref{sub_basic} describe
three variables $P$,$V$ and $u$. In order to obtain self-similarity, we must also assume that the density diverges at the front of the
solution (for further discussion see~\cite{ts}), which makes the solution applicable only for the ablation region.

By defining a dimensionless parameter $\xi$~\cite{ts}:
\begin{equation} \label{super_dim_less}
\xi=m\left( B^{\mu-2}T_0^{2\beta-2\alpha-8-\beta\lambda+(4+\alpha)\mu}t^{-2-2(4+\alpha-\beta)\uptau+\mu(3+(4+\alpha)\uptau)-\lambda(2+\beta\uptau)} \right)^{\frac{1}{4+2\lambda-4\mu}}
\end{equation}
with $B=\frac{16\sigma}{3(4+\alpha)}g( rf)^{\frac{-4+\alpha}{\beta}}$,
the equations of motion, Eqs. \ref{sub_basic}, can be re-written in a self-similar form:
\begin{subequations} \label{sub_ss}
\begin{equation}
\left(w_{V3}+w_3\xi\p{}{\xi}\right)\ti{V}-\p{\ti{u}}{\xi}=0
\end{equation}
\begin{equation}
\left(w_{u3}+w_3\xi\p{}{\xi}\right)\ti{u}+\p{\ti{P}}{\xi}=0
\end{equation}
\begin{align}
& \frac{1}{r}\left[(w_{V3}+w_{P3})\ti{P}\ti{V}+w_3\xi\left(\ti{V}\p{\ti{P}}{\xi}+\ti{P}\p{\ti{V}}{\xi}\right)\right]+\ti{P}\left(w_{V3}+w_3\xi\p{}{\xi}\right)\ti{V}= \\
& \frac{4+\alpha}{\beta} \left\{ \lambda\ti{V}^{\lambda-1}\p{\ti{V}}{\xi}\left(\ti{P}\ti{V}^{1-\mu}\right)^{\frac{4+\alpha-\beta}{\beta}}\left[\ti{P}(1-\mu)\ti{V}^{-\mu}\p{\ti{V}}{\xi}+
\ti{V}^{1-\mu}\p{\ti{P}}{\xi}\right]+\right. \nonumber \\
& \frac{4+\alpha-\beta}{\beta}\ti{V}^{\lambda}\left(\ti{P}\ti{V}^{1-\mu}\right)^{\frac{4+\alpha-2\beta}{\beta}}\left[\ti{P}(1-\mu)\ti{V}^{-\mu}\p{\ti{V}}{\xi}+
\ti{V}^{1-\mu}\p{\ti{P}}{\xi}\right]^2 + \nonumber \\
& \ti{V}^{\lambda}\left(\ti{P}\ti{V}^{1-\mu}\right)^{\frac{4+\alpha-\beta}{\beta}}\left[2(1-\mu)\ti{V}^{-\mu}\p{\ti{V}}{\xi}\p{\ti{P}}{\xi}-
\mu(1-\mu)\ti{P}\ti{V}^{-\mu-1}\left(\p{\ti{V}}{\xi}\right)^2+\right.  \nonumber  \\
&  \left.\left.(1-\mu)\ti{P}\ti{V}^{-\mu}\pp{\ti{V}}{\xi}{2}+\ti{V}^{1-\mu}\pp{\ti{P}}{\xi}{2}\right]\right\} \nonumber
\end{align}
\end{subequations}
Here, the dimensionless variables are written as a function of the dimensional parameters $X=\ti{X}B^{w_{X_1}}T_0^{w_{X_2}}t^{w_{X_3}}$.
The exact values for the powers $w_{X_i}$ for each variable appear in~\cite{ts}.

Simple integration of the self-similar ordinary differential equations (ODE) using the appropriate boundary condition and a multidimensional shooting method, yields the full solution of the ablation region.
For example, for the case of constant temperature boundary condition ($\uptau=0$), the position of the heat-wave front and the ablation pressure can be expressed as a function of $T_0$ and $t$:
\begin{subequations}
\label{sub_T0_quant}
\begin{equation}
m_F=10.17\cdot 10^{-4}T_0^{1.91}t^{0.52}  \left[\mathrm{\frac{g}{cm^2}}\right]
\end{equation}
\begin{equation}
P_F=2.71T_0^{2.63}t^{-0.45}  \left[\mathrm{Mbar}\right]
\label{sub_T0_quant2}
\end{equation}
\end{subequations}

\subsection{The shock region}

In the shock region, radiative heat conduction is negligible, and the hydrodynamics equations purely govern the evolution. These equations are identical to Eqs. \ref{sub_basic} only without the diffusion term in Eq. \ref{sub_energy} which becomes:
\begin{equation}
\label{shock_basic}
\p{e}{t}+P\p{V}{t}=0
\end{equation}

We assume that the boundary condition (between the ablation front and the shock back) is of the form:
\begin{equation}
P(t)=P_0t^{\uptau_S}
\end{equation}
where $P_0\equiv P_F^0 T_0^{\frac{4+\alpha+\beta\lambda-(4+\alpha)\mu}{2+\lambda-2\mu}}$ and $\uptau_S\equiv \frac{-1+\mu+(4+\alpha+\beta\lambda)\uptau-(4+\alpha)\mu\uptau}{2+\lambda-2\mu}$.
$P_F^0$ is the time-independent constant value of the ablation pressure, i.e. for $t=1$nsec and $T_0=100$eV. For example, for the case of constant temperature boundary condition at the ablation region, $P_F^0=2.71$Mbar (see Eq. \ref{sub_T0_quant2}) is yielded from the self-similar solution and serves as a boundary condition for the shock.

On the other boundary (the shock front) Hugoniot relations for strong shock are satisfied~\cite{zeldovich}:
\begin{subequations} \label{hugoniot}
\begin{equation}
\frac{D-u_S}{V_S}=\frac{D}{V_0}
\end{equation}
\begin{equation}
P_S=\frac{Du_S}{V_0}
\end{equation}
\begin{equation}
P_Su_SV_0=D\left(e_S+\frac{u_S^2}{2}\right)
\end{equation}
\end{subequations}
Here, $V_0$ is the unperturbed specific volume, $V_S$, $u_S$, $P_S$, and $e_S$ are the specific volume, matter velocity, pressure, and thermal energy at the shock downstream, and $D$ is the shock velocity in the lab frame.

In the shock region, the EOS differs from the EOS in the ablation region and is assumed to be of an ideal gas with parameter $r_2$: 
\begin{equation}
P(\rho,T)=r_2\rho e(\rho,T)\equiv(\gamma_2-1)\rho e(\rho,T)
\label{def_r2}
\end{equation}

Similarly to the ablation solution, we define a dimensionless parameter $\xi$, which for this case has this form:
\begin{equation}
\label{shock_xi}
\xi=m\left(P_0^{-\nicefrac{1}{2}}V_0^{\nicefrac{1}{2}}t^{-1-\frac{\uptau_S}{2}} \right)
\end{equation}

Using Eqs. \ref{shock_xi} and \ref{def_r2}, the self-similar equations of motion (Eqs. \ref{shock_basic}) are:
\begin{subequations} \label{shock_ss}
\begin{equation}
-\left(1+\frac{\uptau_S}{2}\right)\xi\p{\ti{V}}{\xi}-\p{\ti{u}}{\xi}=0
\end{equation}
\begin{equation}
-\left(1+\frac{\uptau_S}{2}\right)\xi\p{\ti{u}}{\xi}+\frac{\uptau_S}{2}\ti{u}+\p{\ti{P}}{\xi}=0
\end{equation}
\begin{equation}
-\left(1+\frac{\uptau_S}{2}\right)\frac{\xi}{r_2}\ti{V}\p{\ti{P}}{\xi}+\frac{\ti{V}\ti{P}\uptau_S}{r_2}-\left(1+\frac{1}{r_2}\right)\left(1+\frac{\uptau_S}{2}\right)\xi\ti{P}\p{\ti{V}}{\xi}=0.
\end{equation}
\end{subequations}

Here again, the full solution of the shock region is obtained by integrating the self-similar ODEs. For the case of constant temperature ablation region boundary condition, in which the ablation pressure temporal behavior yields $\uptau_S=-0.45$, the shock front position can be expressed as:
\begin{equation}
m_S=7.34\cdot 10^{-3}P_0^{0.5}t^{0.7765}  \left[\mathrm{\frac{g}{cm^2}}\right]
\end{equation}

\begin{figure}
\centering{
\includegraphics*[width=15cm]{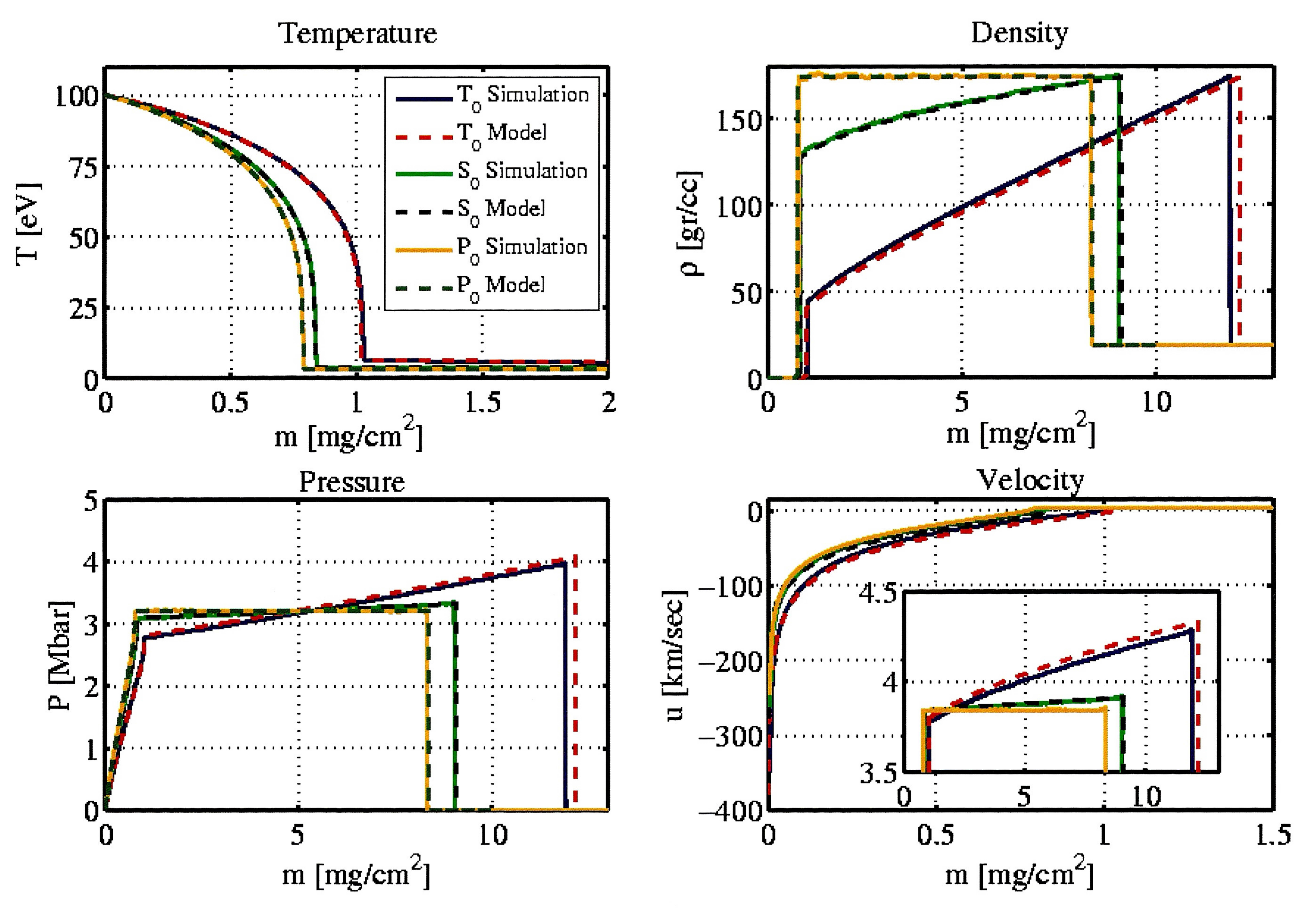}
}
\caption{(Color online) A comparison between the model (dashed lines) and simulations (full lines), under the different 
boundary conditions in late times for a single $r$-value EOS, $r_1=r_2=0.25$. Presented are the profiles of the temperature, density, pressure and matter velocity. The small box in the velocity
graph shows a close-up of the shock region. The figure is taken from~\cite{ts}.}
\label{late}
\end{figure}

In Fig. \ref{late} we present a comparison between the model described above, and full 1D radiation-hydrodynamics numerical simulations for
a single $r$-value EOS ($r_1=r_2=0.25$) and different boundary conditions at $t=1$nsec. The match between all profiles in the
ablation region is within $1\%$, and in the shock region is within $5\%$.

\section{Statement of the problem}
\label{motivation}

In a previous work (also showed in Fig. \ref{late})~\cite{ts}, a single EOS was used for both regions ($r_1=r_2=0.25$). We remind that this value is a fit
for the region $T=100-300\mathrm{eV}$ (ablation region temperatures). However, the best fit for an ideal gas EOS for shock region temperatures ($T\approx 1-10\mathrm{eV}$) is extremely different.
As a result, the model may not describe correctly the solution at the shock region. 

\begin{figure}
\centering{
\includegraphics*[width=8cm]{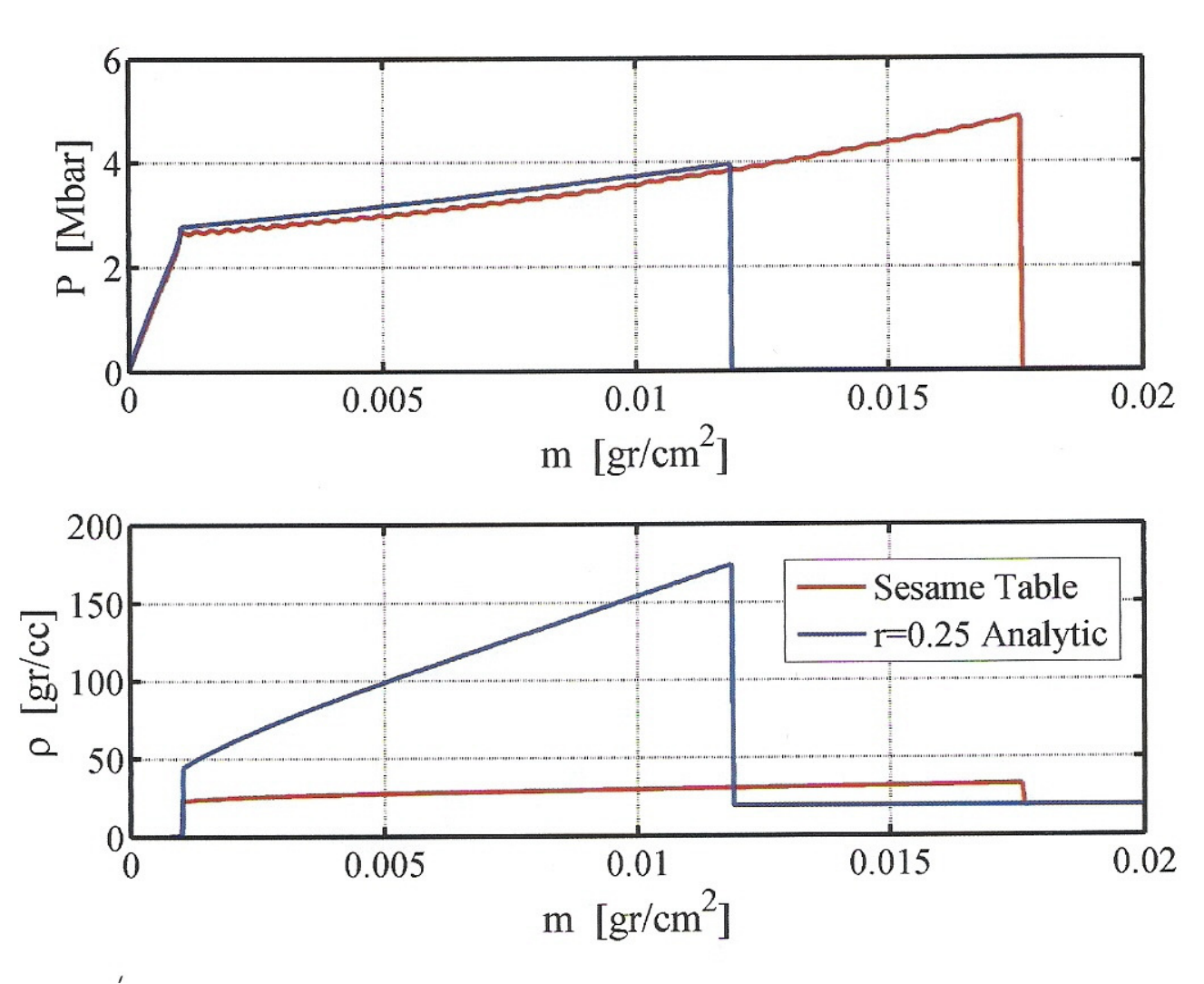}
}
\caption{(Color online) The pressure and density profiles, obtained from one-dimensional numerical calculations with $T=100eV$ constant temperature boundary condition after $t=1$nsec. Compared are the results of a full EOS SESAME table~\cite{sesame} and an ideal gas EOS with numerical value $r=\gamma-1=0.25$ (proposed by Hammer \& Rosen~\cite{hr}).}
\label{motiv1}
\end{figure}

\begin{figure}
\centering{
\includegraphics*[width=8cm]{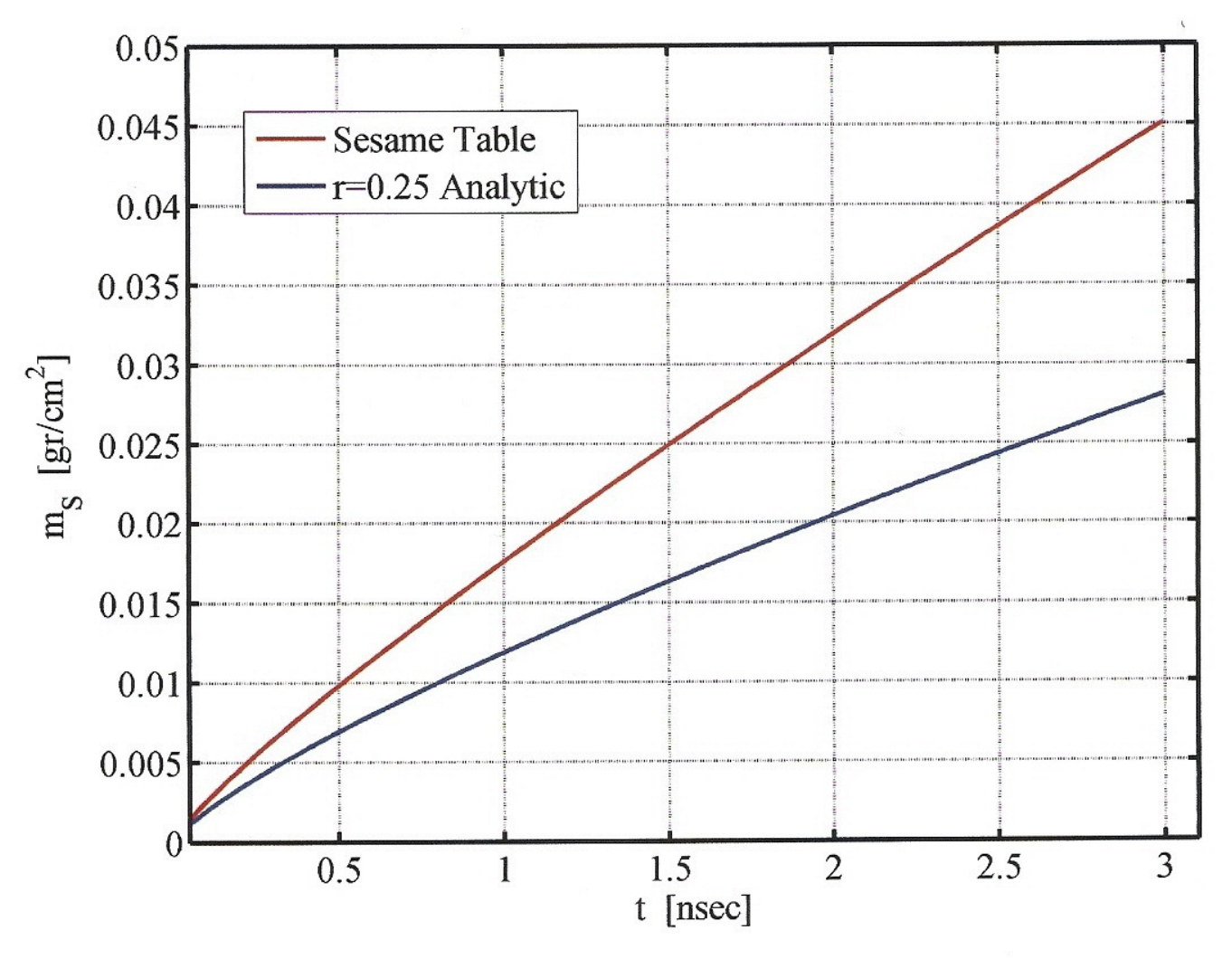}
}
\caption{(Color online) The shock front position $m_S$ as a function of time obtained from one-dimensional numerical calculations with $T=100eV$ constant temperature boundary condition. Compared are the results of a full EOS SESAME table~\cite{sesame} and an ideal gas EOS with numerical value $r=\gamma-1=0.25$ (proposed by Hammer \& Rosen~\cite{hr}).}
\label{motiv2}
\end{figure}

In Figs. \ref{motiv1} and \ref{motiv2} we present the results of two numerical simulations, one with a single ideal gas EOS $r_1=r_2=0.25$ and
one with a SESAME table EOS~\cite{sesame} instead of the simple power-law EOS (Eqs. \ref{pwrlaw_energy}, \ref{def_r} and \ref{def_r2}). In both simulations,
a power-law shape of the Rossland mean opacity, Eq. \ref{ross}, is still assumed (The opacity is in fact relevant only in the ablation region).
In Fig. \ref{motiv1} the pressure and density profiles are shown at $t=1$nsec for a constant temperature boundary condition of
$T_0=100$eV. The shock wave position $m_S$ as a function of the time is plotted for the same boundary condition in Fig. \ref{motiv2}.

We can see that although the calculations with ideal gas EOS fit the SESAME calculations well in the ablation region (within $1\%$), they fail to reproduce the results in the shock region.
The pressure profile obtained by the model fits the table calculations well, until the shock front, which is not well reproduced itself.
The density profile obtained by the model is completely different from the table obtained profile. In particular, the maximal density (which is in the strong shock limit equals
$\rho_0(r_2+2)/r_2=\rho_0(\gamma_2+1)/(\gamma_2-1)$), reaches $\rho_{\max}\approx175$gr/cc in the model, while the table calculated maximal density reaches $\rho_{\max}\approx35-40$gr/cc.
As a result of energy conservation, the shock front position is also different between the simulation and the model, as the table calculated $\dot{m}_S$ is about twice faster than in the constant-$r$ EOS calculation.

These results emphasize the importance of using a correct EOS in each physical (temperature and density) regime of the problem. Since our solution is composed of two different solutions (which are patched together),
it is straightforward to try and use a ``binary-$r$" EOS, where the ideal gas parameter in the ablation region ($r_1$) has a different value than the ideal gas parameter in the shock region ($r_2$).
Even when using a binary-$r$ model, where each value of $r$ is fitted to a table for its own regime, it is not straightforward that the model will reproduce the results obtained by a full numerical calculation using the table itself.

In the following section we demonstrate our model, and show that for heat waves in Au and boundary temperatures $100-300\rm{eV}$, the model can be used to obtain a good approximate solution which is valid on also in the shock region. This work
also shows the benefit of using a solution which is composed of two different solutions for two physically different regions, as opposed to~\cite{garnier}.

\section{Binary-$r$ equation of state}
\label{results}

In the $T=100-300$eV regime, the EOS can fits reasonably to ideal gas with a single $r_1=0.25$. On the other hand, the EOS for shock region temperatures 
($T=1-30eV$, highly dependent on the boundary conditions in the ablation region), varies significantly. 
Thus, restricted to a single $r$-value of the EOS in shock regime, $r_2$, one must fit the value
for a narrow thermodynamic range of the EOS. The specific boundary condition sets the temperature criterion, which sets the relevant temperature range of the EOS.

\begin{figure}
\centering{
(a)
\includegraphics*[width=7.4cm]{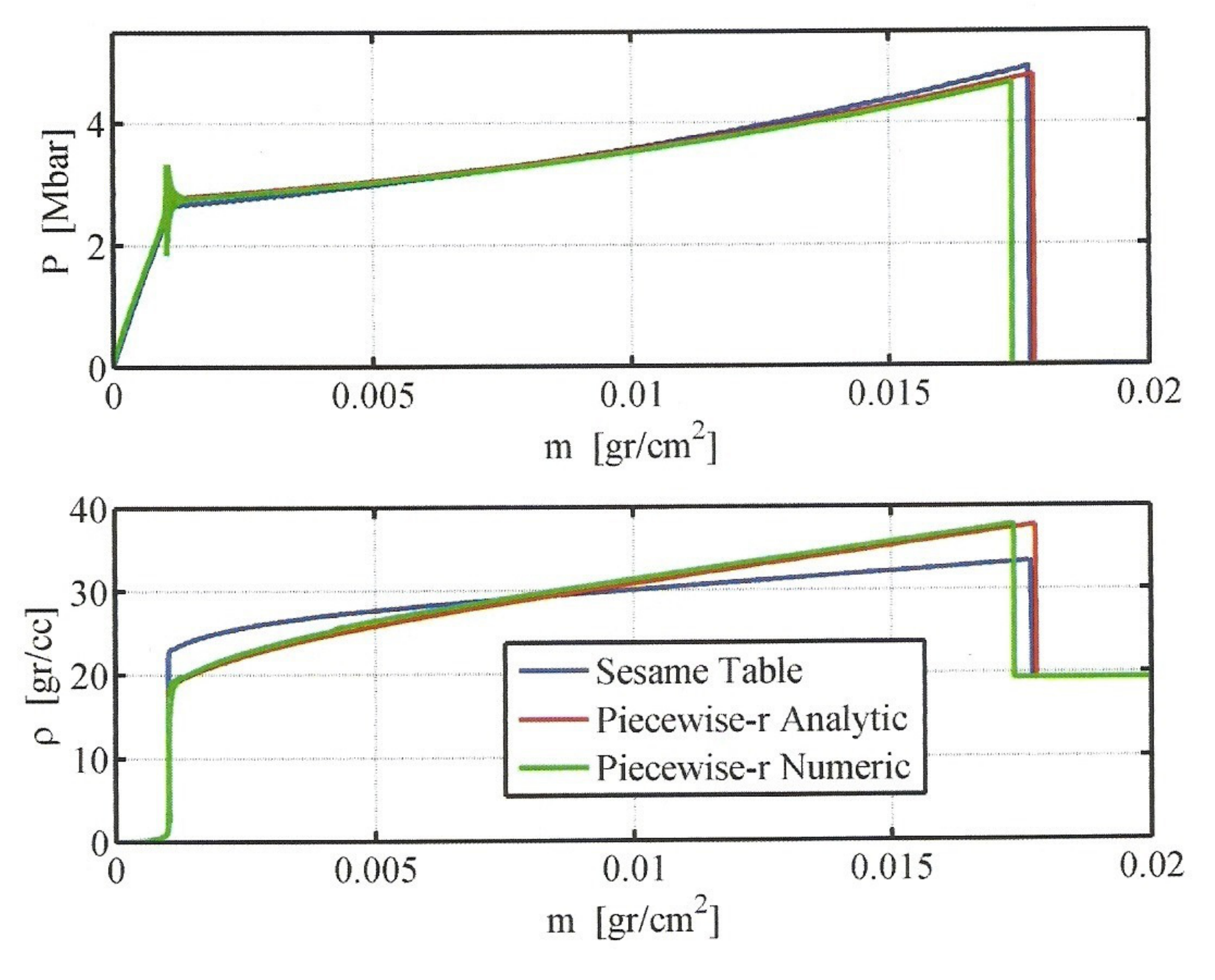}
(b)
\includegraphics*[width=7.5cm]{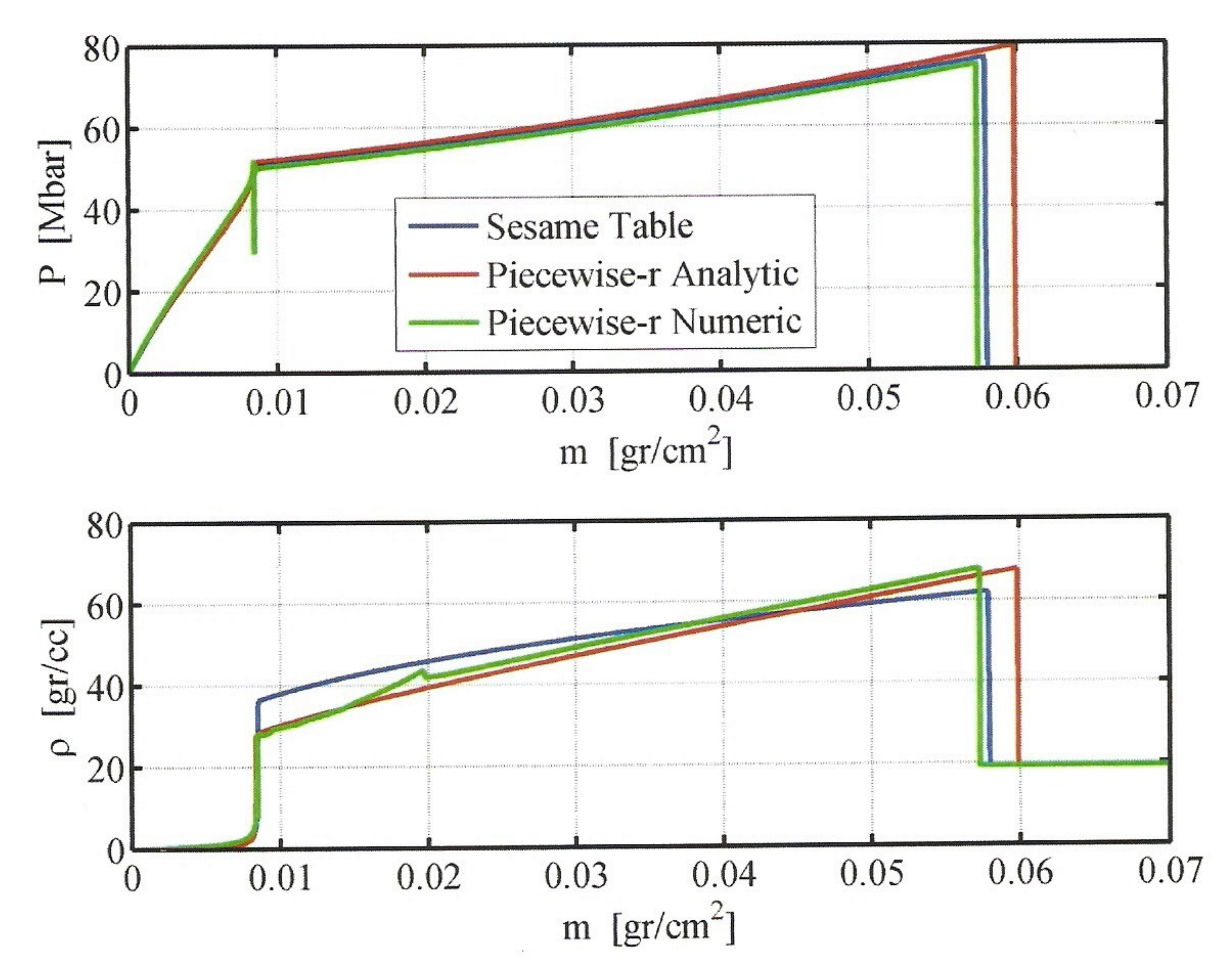}
}
\caption{(Color online) (a) The pressure and density profiles using a binary-EOS model, a binary-EOS numerical simulation and a SESAME table EOS numerical simulation,
for a constant temperature boundary condition $T_S=T_0=100$eV and in a representative time, $t=1$nsec. (b) The same for {\em higher} boundary temperature $T_S=T_0=300$eV.}
\label{p_rho}
\end{figure}

\begin{figure}
\centering{
(a)
\includegraphics*[width=7.9cm]{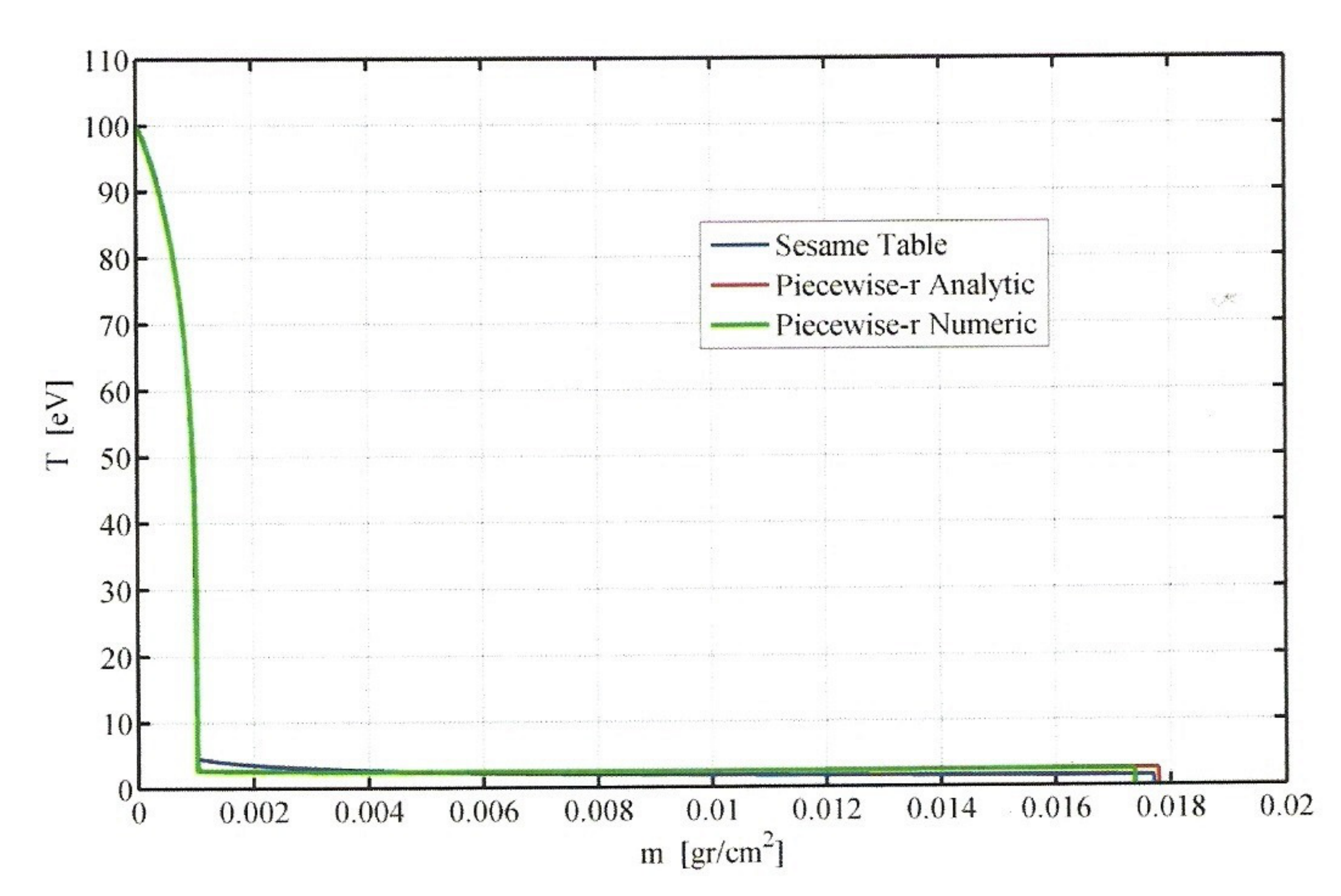}
(b)
\includegraphics*[width=6.5cm]{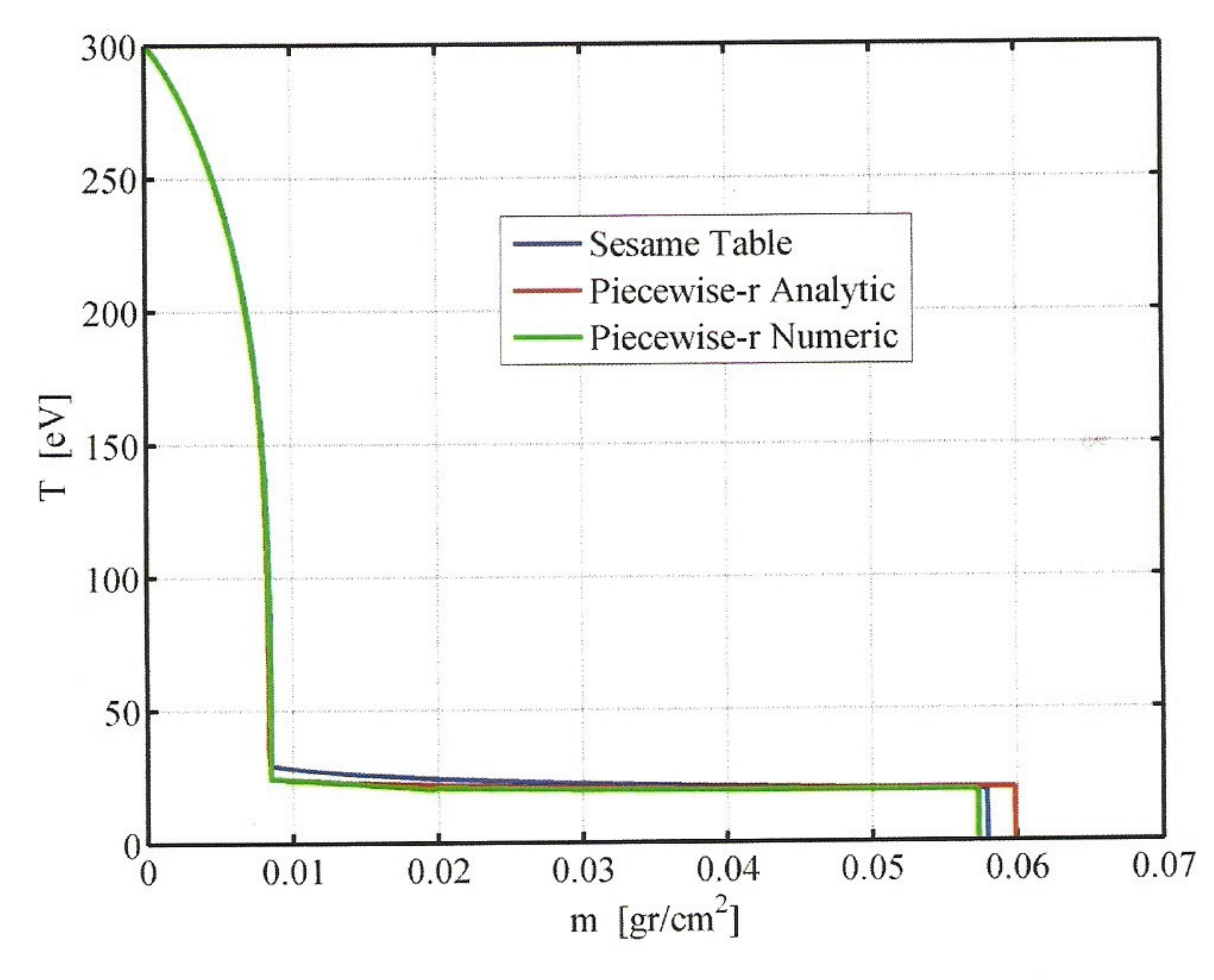}
}
\caption{(Color online) (a) The temperature profile using a binary-EOS model, a binary-EOS numerical simulation and a SESAME table EOS numerical simulation, for a constant temperature boundary condition $T_S=T_0=100$eV and in a representative time, $t=1$nsec.
(b) The same for {\em higher} boundary $T_S=T_0=300$eV.}
\label{temp}
\end{figure}

For the case of boundary condition with $T_0=100$eV, the temperature in the shock region is about $2-10$eV. For these temperatures, the best fit value of the EOS to an ideal gas is $r_2=2.1$
(with a large variance of$\approx50\%$, due to the highly-dependency of the EOS in the shock regime, that may cause errors
in describing the full physical behavior). Thus, we shall use the self-similar model with $r_1=0.25$ in the ablation region and $r_2=2.1$ in the shock region.
The model's results are presented in Figs. \ref{p_rho}(a)-\ref{temp}(a) and compared to numerical simulations using both binary-$r$ EOS and SESAME table EOS.
In addition, we present the results of a 1D numerical simulation, with a binary-$r$ EOS that is defined as:
\begin{equation}
\label{r_bin1}
r=\begin{cases} 0.25 \quad & T>10\mathrm{eV} \\
2.1 & T\leqslant10\mathrm{eV}
\end{cases}
\end{equation}

As can be seen in Fig. \ref{p_rho}(a), the model and the binary-$r$ simulations fit to within $3\%$ accuracy. This result is an extension of the results from~\cite{ts}, now for binary-$r$ EOS.
In addition, the binary-$r$ simulation and the SESAME table calculations fit to within $2\%$ in the pressure profile and in the front coordinate (we note that
the match between the binary-$r$ semi-analytic model and the SESAME table calculations is better because these two errors reduce each other). As for the density profile,
both the model and the binary-$r$ calculation reproduce the table results to within $15\%$. The lower agreement is because the density is most sensitive to the numerical
value of the parameter in ideal gas EOS. Thus, the binary-$r$ model predicts the solution of the shock region much better than a single EOS model (See again Fig. \ref{motiv1}).
The temperature profile for the same boundary condition are presented in Fig. \ref{temp}(a). In the ablation region, the agreement
is within $1\%$, while in the shock region, the agreement is within $15\%$, which is of course similar to the agreement of the density profile.

For higher ablation temperatures, higher shock temperatures are also obtained. For example, in the temperature regime of NIF experiments~\cite{lindl} ($T_0\approx300$eV) the expected
shock temperature is of about $20-30$eV. For these higher temperatures, SESAME tables (and EOS in general) fit much better to ideal gas EOS because the high internal energy is larger
compared to solid lattice interactions. We therefore expect the binary-$r$ model to better predict the full simulation results in higher temperatures.
A best fit of the EOS for temperatures of $20-30$eV, yields $r_2=0.8$, and thus the appropriate EOS to be used in the binary-$r$ simulation is:
\begin{equation}
\label{r_bin2}
r=\begin{cases} 0.25 \quad & T>50\mathrm{eV} \\
0.8 & T\leqslant50\mathrm{eV}
\end{cases}
\end{equation}

In Figs. \ref{p_rho}(b)-\ref{temp}(b) we present a comparison between the semi-analytic model and numerical simulations, for conditions similar
to Figs. \ref{p_rho}(a)-\ref{temp}(a), only now with a temperature $T_S=T_0=300$eV. Here again, the agreement between the model and the binary-$r$
simulation within $3\%$ everywhere (except the difference at the shock position between the semi-analytic model and the numerical
simulation; for details on this error see~\cite{ts}). As for the comparison with the SESAME table EOS, we can see the binary-$r$ simulation fit the SESAME table simulation
better for the higher temperature (within $10\%$ for the density profile).

\begin{figure}
\centering{
(a)
\includegraphics*[width=7.5cm]{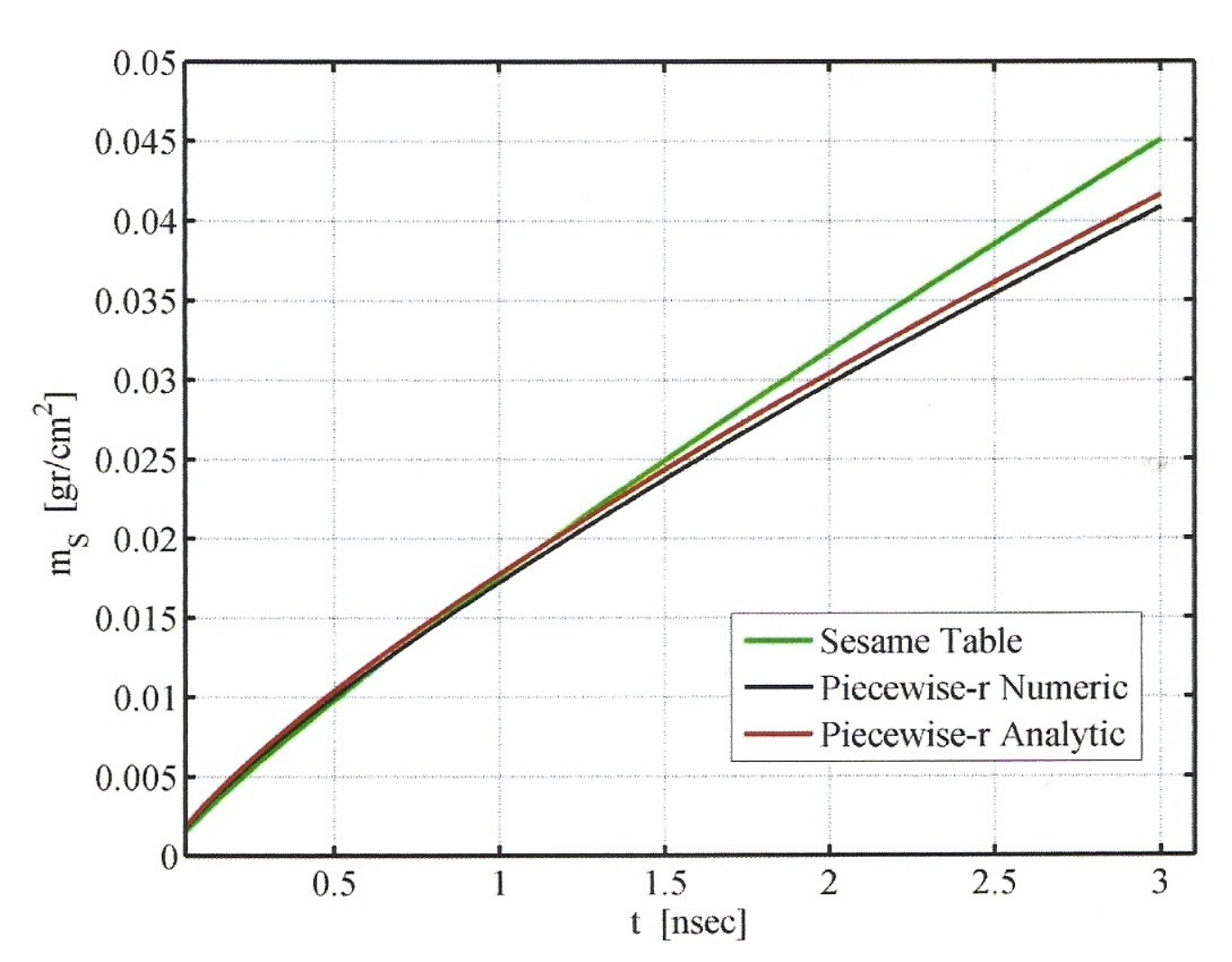}
(b)
\includegraphics*[width=7.3cm]{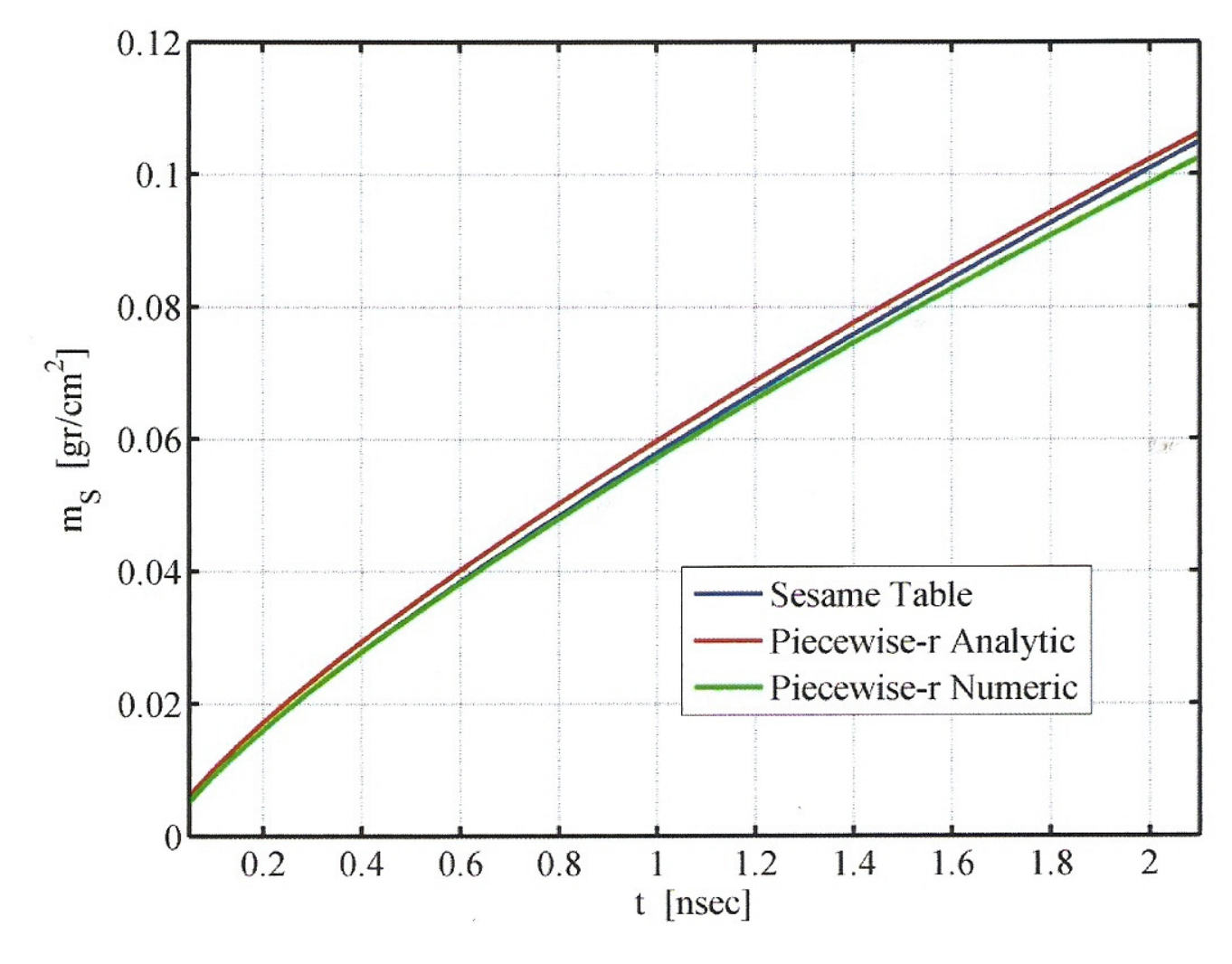}
}
\caption{(Color online) (a) The shock front $m_S$ as a function of time, using a binary-EOS model, a binary-EOS numerical simulation and a SESAME table EOS numerical simulation, for a constant temperature boundary condition $T_S=T_0=100$eV.
(b) The same for {\em higher} constant boundary temperature $T_S=T_0=300$eV.}
\label{shock_front}
\end{figure}

It is important to note that because the temperatures and densities of the shock region change with time, also the best fit value of $r_2$ changes.
Therefore, comparing the hydrodynamic profiles obtained by the model at a single time $t=1$nsec is not enough.
For testing our model as a function of the time, in Fig. \ref{shock_front} we present the shock front $m_S$ as a function of time for different $T_0$. It can be seen that
the self-similar model yields the correct time dependency of the shock front position for all the times. In $T_0=100$eV, the maximal deviation is about $10\%$, while
in $T_0=300$eV, the maximal deviation is lower (about $5\%$), as expected and explained before.
Comparing these values of errors to
the single $r$ value (Fig. \ref{motiv2}) of factor of 2 in the shock front, emphasis the use of binary-$r$ self-similar solutions.

\section{Discussion}
\label{discussion}

In this work, we used the fact that the semi-analytic solution of~\cite{ts} is composed of two separate self-similar solutions, each one valid in a different
region of the physical problem, in order to use a different EOS for each region while maintaining the self-similarity of the problem. This model was tested for Au thin foils.
We showed that using the naive approximation that the EOS is the same in both regions of a single $r$-value of EOS causes large errors in describing the
shock region (compared to a similar calculation with a SESAME table), both in the hydrodynamic
profiles and in and the shock position. Using a binary-$r$ EOS where the value $r_2$ in the shock region is best fit to SESAME table, yields a much better prediction of the physical
solution of the shock region. We checked the model for several temperatures, all in the relevant range of HEDP experiments, and found that the model is applicable for designing,
analyzing and understanding most HEDP experiments. The model is especially useful for understanding and designing experiments in which the shock is important, such as
EOS experiments or subsonic heat wave experiments. 

\begin{acknowledgments}
The authors are grateful to Nir Sapir for useful discussions.
\end{acknowledgments}

\end{document}